\documentclass[12pt]{article}
\usepackage[cp1251]{inputenc}

\newcommand{\Tr}{\mathop{\rm Tr}\nolimits}

\title{About calculation of traces of Dirac $\gamma$-matrices
              contracted with massless vectors in Minkowski space}

\author {Alexander~L.~Bondarev$^1$ and Alexander~R.~Roslik$^2$
 \\ \\ \it \normalsize National Scientific and Educational Center of
 Particle and High Energy Physics
 \\ \it \normalsize of the Belarusian State University \\
    \it \normalsize M.Bogdanovich str.,153, Minsk 220050,
      Belarus \\
  \rm \normalsize  ${}^1$e-mail: bondarev@hep.by \\
  \rm \normalsize  ${}^2$e-mail: roslik@hep.by }

\date{ }

\begin{document}
\maketitle

\begin{abstract}
A new method for calculation of traces of Dirac $\gamma$-matrices
contracted with massless vectors in Minkowski space is discussed
\end{abstract}

\section{Introduction}

Calculation of Dirac $\gamma$-matrices traces is standard procedure
in high energy physics
computation. However the expressions for the traces of more than ten
$\gamma$-matrices are too long.

Application of Chisholm \cite{r1} identities and Kahane \cite{r2}
algorithm leads to short expressions for traces in 4-spacetime.
More compact formulae can be obtained through the new method
proposed in \cite{r3} and based on the properties of orthonormal
bases in the Minkowski space and isotropic tetrads (see e.g.
\cite{r4}-\cite{r5}) constructed from the vectors of these bases.

The purpose of the article is further simplification of the formulae
presented in \cite{r3} for the case of Dirac $\gamma$-matrices
contracted with massless vectors. This task is important today because
most of analytical calculations in high energy physics are performed at
massless approximation (see, for example, \cite{r6}).

The article is arranged in the  following way. The new method of
trace computation for arbitrary vectors is shortly described in
Section 2. In Section 3 one explores application of the traces
calculation method for Dirac $\gamma$-matrices contracted with
massless vectors. The formulae obtained there are good enough to
realize an algorithm for any numerical calculation.

Note, that the method of calculation of the traces from the
\cite{r3} is included in the system of analytical computation
ALHEP \cite{r8}, and this article contains the proposals for
further improvement of the similar programs.


In this paper we use the Feynman metrics:
$$
\begin{array}{l} \displaystyle
\mu = 0,1,2,3, \;\;\; a^{\mu} = ( a_0 , \vec{a} ) ,
 \;\;\;
 a_{\mu} = ( a_0, -\vec{a} ) ,
 \;\;\;
(a b) = a_{\mu} b^{\mu}
= a_0 b_0 - \vec{a} \vec{b} \; ,
\end{array}
$$
and a sign of the Levi-Civita tensor is fixed as
$$ \displaystyle \;
{\varepsilon}_{ 0 1 2 3 } = + 1 \, .$$

Orientation of the orthonormal basis vectors in Minkowski space
$$\displaystyle (l_A l_B) =  g_{A B} \; , \; \; (A,B = 0,1,2,3) $$
is constrained by
$$\displaystyle
\varepsilon_{\mu \nu \rho \lambda} l_0^{\mu} l_1^{\nu} l_2^{\rho}
l_3^{\lambda} = + 1  \; .
$$

Isotropic tetrads $q_{\pm}$, $e_{\pm}$ are constructing in the following way:
\begin{equation}
\begin{array}{l} \displaystyle
 q_{+} = { l_0 + l_1 \over \sqrt{2} } \; , \;\;
 q_{-} = { l_0 - l_1 \over \sqrt{2} } \; , \;\;
 (q_{+} q_{-}) = 1 \; ;
          \\[0.5cm] \displaystyle
 e_{+} = { l_2 + i l_3 \over \sqrt{2} } \; , \;\;
 e_{-} = { l_2 - i l_3 \over \sqrt{2} } \; , \;\;
 (e_{+} e_{-}) = -1 \; , \;\;
 e_{\mp}^{\ast} = e_{\pm} \; ;
          \\[0.5cm] \displaystyle
(q_{+} e_{+}) = (q_{+} e_{-}) = (q_{-} e_{+}) = (q_{-} e_{-}) = 0 \; .
\end{array}
\label{e.1-1}
\end{equation}
%


\section{The new method of trace calculation for Dirac $\gamma$-matrices
contracted with arbitrary  vectors}

The new method to calculate traces of Dirac $\gamma$-matrices
contracted with arbitrary vectors $a_i$ $(i =1, \cdots, 2n)$ in
Minkowski space is presented in \cite{r3}. So the following
expressions were obtained:
\begin{equation}
{1 \over 2} \Tr [(1 - \gamma^5) \hat{a}_1 \hat{a}_2 ]
= F_1(a_1, a_2) + F_3(a_1, a_2)
\; ;
\label{e.1-2}
\end{equation}
\begin{equation}
\begin{array}{l} \displaystyle
{1 \over 2} \Tr [(1 - \gamma^5) \hat{a}_1 \hat{a}_2 \hat{a}_3
\hat{a}_4 ] =
          \\[0.5cm] \displaystyle
 = F_1(a_1, a_2) F_1(a_3, a_4) + F_2(a_1, a_2) F_4(a_3, a_4)
+ F_3(a_1, a_2) F_3(a_3, a_4) + F_4(a_1, a_2) F_2(a_3, a_4)
 \; ;
\end{array}
\label{e.1-3}
\end{equation}
where $F_1$, $F_2$, $F_3$, $F_4$ are some functions (see below) of
vectors $a_i$.

Hawing calculated
$$ \displaystyle \Tr [(1 - \gamma^5) \hat{a}_1 \cdots
\hat{a}_{2n-1} \hat{a}_{2n} ] \; , $$
we may obtain an expression for
$$ \displaystyle \Tr [(1 - \gamma^5) \hat{a}_1 \cdots
\hat{a}_{2n-1} \hat{a}_{2n} \hat{a}_{2n+1} \hat{a}_{2n+2} ] $$
from the previous one by the following replacement:
\begin{equation}
\begin{array}{r} \displaystyle
F_1 (a_{2n-1}, a_{2n})
 \rightarrow F_1 (a_{2n-1}, a_{2n}) F_1 (a_{2n+1}, a_{2n+2})
 + F_2 (a_{2n-1}, a_{2n}) F_4 (a_{2n+1}, a_{2n+2}) \; ,
          \\[0.5cm] \displaystyle
F_3 (a_{2n-1}, a_{2n})
 \rightarrow F_3 (a_{2n-1}, a_{2n}) F_3 (a_{2n+1}, a_{2n+2})
 + F_4 (a_{2n-1}, a_{2n}) F_2 (a_{2n+1}, a_{2n+2}) \; ,
          \\[0.5cm] \displaystyle
F_2 (a_{2n-1}, a_{2n})
 \rightarrow F_1 (a_{2n-1}, a_{2n}) F_2 (a_{2n+1}, a_{2n+2})
 + F_2 (a_{2n-1}, a_{2n}) F_3 (a_{2n+1}, a_{2n+2}) \; ,
          \\[0.5cm] \displaystyle
F_4 (a_{2n-1}, a_{2n})
 \rightarrow F_3 (a_{2n-1}, a_{2n}) F_4 (a_{2n+1}, a_{2n+2})
 + F_4 (a_{2n-1}, a_{2n}) F_1 (a_{2n+1}, a_{2n+2})
\; .
\end{array}
\label{e.1-4}
\end{equation}

The expression for
$$ \displaystyle
 \Tr [(1 - \gamma^5) \hat{a}_1 \cdots \hat{a}_{2n-1} \hat{a}_{2n} ] $$
obtained through the method will contain $2^n$ terms.

\begin{equation}
\Tr [(1 + \gamma^5) \hat{a}_1 \hat{a}_2 \cdots \hat{a}_{2n} ] =
\left( \Tr [(1 - \gamma^5) \hat{a}_1 \hat{a}_2 \cdots \hat{a}_{2n} ] \right)^{\ast} \; .
\label{e.1-5}
\end{equation}

The functions mentioned above have the following forms:
\begin{equation}
\begin{array}{l} \displaystyle
F_1 (a_i, a_j) =
 2 [ (a_i q_{-}) (a_j q_{+}) - (a_i e_{+}) (a_j e_{-}) ] =
          \\[0.3cm] \displaystyle
=  (a_i a_j) +
\begin{array}{l}
 G\pmatrix{ a_i & a_j \\ l_0 & l_1 }
\end{array}
  + i
\begin{array}{l}
 G\pmatrix{ a_i & a_j \\ l_2 & l_3 }
\end{array} =
          \\[0.5cm] \displaystyle
 =  {1 \over 4} \Tr [ (1 - \gamma^5) \hat{q}_{+} \hat{q}_{-} \hat{a}_i
\hat{a}_j ]
 = - {1 \over 4} \Tr [ (1 - \gamma^5) \hat{e}_{-} \hat{e}_{+} \hat{a}_i
\hat{a}_j ] \; ,
\end{array}
\label{e.1-6}
\end{equation}
\begin{equation}
\begin{array}{l} \displaystyle
F_3 (a_i, a_j) =
 2 [ (a_i q_{+}) (a_j q_{-}) - (a_i e_{-}) (a_j e_{+}) ] =
          \\[0.3cm] \displaystyle
= (a_i a_j) -
\begin{array}{l}
 G\pmatrix{ a_i & a_j \\ l_0 & l_1 }
\end{array}
- i
\begin{array}{l}
 G\pmatrix{ a_i & a_j \\ l_2 & l_3 }
\end{array} =
          \\[0.5cm] \displaystyle
= {1 \over 4} \Tr [ (1 - \gamma^5) \hat{q}_{-} \hat{q}_{+} \hat{a}_i
\hat{a}_j ]
 = - {1 \over 4} \Tr [ (1 - \gamma^5) \hat{e}_{+} \hat{e}_{-} \hat{a}_i
\hat{a}_j ] \; ,
\end{array}
\label{e.1-7}
\end{equation}
\begin{equation}
\begin{array}{l} \displaystyle
F_2 (a_i, a_j) = 2 [ (a_i e_{+}) (a_j q_{-}) - (a_i q_{-}) (a_j e_{+}) ]
 = 2
\begin{array}{l}
 G\pmatrix{ a_i & a_j \\ e_{+} & q_{-} }
\end{array} =
          \\[0.5cm] \displaystyle
= -
\begin{array}{l}
 G\pmatrix{ a_i & a_j \\ l_0 & l_2 }
\end{array}
  + i
\begin{array}{l}
  G\pmatrix{ a_i & a_j \\ l_1 & l_3 }
\end{array}
 +
\begin{array}{l}
 G\pmatrix{ a_i & a_j \\ l_1 & l_2 }
\end{array}
  - i
\begin{array}{l}
 G\pmatrix{ a_i & a_j \\ l_0 & l_3 }
\end{array} =
          \\[0.5cm] \displaystyle
= {1 \over 4} \Tr [ (1 - \gamma^5) \hat{q}_{-} \hat{e}_{+} \hat{a}_i
\hat{a}_j ] \; ,
\end{array}
\label{e.1-8}
\end{equation}
\begin{equation}
\begin{array}{l} \displaystyle
F_4 (a_i, a_j) =
 2 [ (a_i e_{-}) (a_j q_{+}) - (a_i q_{+}) (a_j e_{-}) ] = 2
\begin{array}{l}
 G\pmatrix{ a_i & a_j \\ e_{-} & q_{+} }
\end{array} =
          \\[0.5cm] \displaystyle
 =  -
\begin{array}{l}
 G\pmatrix{ a_i & a_j \\ l_0 & l_2 }
\end{array}
+ i
\begin{array}{l}
G\pmatrix{ a_i & a_j \\ l_1 & l_3 }
\end{array}
-
\begin{array}{l}
 G\pmatrix{ a_i & a_j \\ l_1 & l_2 }
\end{array}
+ i
\begin{array}{l}
 G\pmatrix{ a_i & a_j \\ l_0 & l_3 }
\end{array} =
           \\[0.5cm] \displaystyle
 =  {1 \over 4} \Tr [ (1 - \gamma^5) \hat{q}_{+} \hat{e}_{-} \hat{a}_i
\hat{a}_j ] \; ,
\end{array}
\label{e.1-9}
\end{equation}
where $G$ are Gram determinants.

There is an example of $F_i$ functions for some orthonormal basis.
Let's to fix the last as follow:
\begin{equation}
\begin{array}{l} \displaystyle
 l^{\mu}_0 = (1,\ 0, 0, 0) \; , \;\;
 l^{\mu}_1 = (0,\ 1, 0, 0) \; , \;\;
 l^{\mu}_2 = (0,\ 0, 1, 0) \; , \;\;
 l^{\mu}_3 = (0,\ 0, 0, 1) \;.
\end{array}
\label{e.1-10}
\end{equation}
Then the isotropic tetrads can be expressed as
\begin{equation}
\begin{array}{l} \displaystyle
q^{\mu}_{\pm} = {1 \over \sqrt{2} } (1,\ \pm 1 , 0, 0)
 \; , \; \;
 (a q_{\pm} ) = {1 \ \over \sqrt{2} } ( a_0 \mp a_x) \; ,
          \\[0.5cm] \displaystyle
 e^{\mu}_{\pm} = {1 \over \sqrt{2} } (0,\ 0 , 1, \pm i)
 \; , \; \;
(a e_{\pm} ) = - {1 \ \over \sqrt{2} } ( a_y \pm i a_z)
 \; .
\end{array}
\label{e.1-11}
\end{equation}
At last
\begin{equation}
\begin{array}{r} \displaystyle
F_1 (a_i, a_j) = (a_i a_j) -
 \left[ \ (a_i)_0 (a_j)_x - (a_i)_x (a_j)_0 \ \right]
 + i \left[ \ (a_i)_y (a_j)_z - (a_i)_z (a_j)_y \ \right] \; ,
          \\[0.5cm] \displaystyle
F_3 (a_i, a_j) = (a_i a_j) +
 \left[ \ (a_i)_0 (a_j)_x - (a_i)_x (a_j)_0 \ \right]
 - i \left[ \ (a_i)_y (a_j)_z - (a_i)_z (a_j)_y \ \right]  \; ,
          \\[0.5cm] \displaystyle
F_2 (a_i, a_j) =
  \left[ \ (a_i)_0 (a_j)_y - (a_i)_y (a_j)_0 \ \right] +
  i \left[ \ (a_i)_x (a_j)_z - (a_i)_z (a_j)_x \ \right] +
          \\[0.3cm] \displaystyle
+ \left[ \ (a_i)_x (a_j)_y - (a_i)_y (a_j)_x \ \right] +
 i \left[ \ (a_i)_0 (a_j)_z - (a_i)_z (a_j)_0 \ \right] \; ,
          \\[0.5cm] \displaystyle
F_4 (a_i, a_j) =
 \left[ \ (a_i)_0 (a_j)_y - (a_i)_y (a_j)_0 \ \right]
+ i  \left[ \ (a_i)_x (a_j)_z - (a_i)_z (a_j)_x \ \right] -
          \\[0.3cm] \displaystyle
-  \left[ \ (a_i)_x (a_j)_y - (a_i)_y (a_j)_x \ \right]
 - i  \left[ \ (a_i)_0 (a_j)_z - (a_i)_z (a_j)_0 \ \right] \; .
\end{array}
\label{e.1-12}
\end{equation}

For clarity one can writes expressions of traces in the following form:
\begin{equation}
\begin{array}{l} \displaystyle
{1 \over 2} \Tr \left[ (1 - \gamma^5){\hat a}_1 {\hat a}_2 {\hat a}_3 {\hat a}_4
 \cdots {\hat a}_{2n-1} {\hat a}_{2n} \right] =
          \\[0.5cm] \displaystyle
 = \Tr \Bigg(
\left[
\begin{array}{rr}
  F_3(a_1, a_2) & F_4(a_1, a_2) \\
  F_2(a_1, a_2) & F_1(a_1, a_2) \\
\end{array}%
\right]
\cdot
\left[
\begin{array}{rr}
  F_3(a_3, a_4) & F_4(a_3, a_4) \\
  F_2(a_3, a_4) & F_1(a_3, a_4) \\
\end{array}%
\right]
 \cdot \, \cdots \, \cdot
\left[
\begin{array}{rr}
  F_3(a_{2n-1}, a_{2n}) & F_4(a_{2n-1}, a_{2n}) \\
  F_2(a_{2n-1}, a_{2n}) & F_1(a_{2n-1}, a_{2n}) \\
\end{array}%
\right]
\Bigg)  \; .
\end{array}
\label{e.1-13}
\end{equation}
It's obvious that the trace of $2n$ Dirac $\gamma$-matrices is reduced  to trace of $n$ matrices with
$2 \times 2$ dimension.

For calculation of expressions like  this
$$ \displaystyle
 \Tr ( {\gamma}_{\rho} \hat{a}_1 \hat{a}_2 \cdots \hat{a}_{2n+1} )
 \Tr ( {\gamma}^{\rho} \hat{b}_1 \hat{b}_2 \cdots \hat{b}_{2m+1} )
\; ,
$$
where summing over index $\rho$ is supposed, one can use the
Fiertz transform:
\begin{equation}
\displaystyle
[(1 \pm {\gamma}^5) {\gamma}_{\rho} ]_{ij}
[(1 \mp {\gamma}^5) {\gamma}_{\rho} ]_{kl}
 = 2 (1 \pm {\gamma}^5)_{il} (1 \mp {\gamma}^5)_{kj} \; .
\label{e.1-14}
\end{equation}

So one achieves the following expressions:
\begin{equation}
\begin{array}{l} \displaystyle
 \Tr [ (1 \pm {\gamma}^5) {\gamma}_{\rho} \hat{a}_1 \hat{a}_2 \cdots \hat{a}_{2n+1} ]
 \cdot \Tr [ (1 \mp {\gamma}^5) {\gamma}^{\rho} \hat{b}_1 \hat{b}_2 \cdots \hat{b}_{2m+1}
 ] =
          \\[0.5cm] \displaystyle
= 4 \Tr [ (1 \mp {\gamma}^5) \hat{a}_1 \hat{a}_2 \cdots \hat{a}_{2n+1}
 \hat{b}_1 \hat{b}_2 \cdots \hat{b}_{2m+1} ] \; ,
\end{array}
\label{e.1-15}
\end{equation}
\begin{equation}
\begin{array}{l} \displaystyle
 \Tr [ (1 \pm {\gamma}^5) {\gamma}_{\rho} \hat{a}_1 \hat{a}_2 \cdots \hat{a}_{2n+1} ]
 \cdot \Tr [ (1 \pm {\gamma}^5) {\gamma}^{\rho} \hat{b}_1 \hat{b}_2 \cdots \hat{b}_{2m+1}
 ] =
          \\[0.5cm] \displaystyle
= \Tr [ (1 \pm {\gamma}^5) {\gamma}_{\rho} \hat{a}_1 \hat{a}_2 \cdots \hat{a}_{2n+1} ]
 \cdot \Tr [ (1 \mp {\gamma}^5) {\gamma}^{\rho} \hat{b}_{2m+1} \cdots  \hat{b}_2 \hat{b}_1
 ] =
          \\[0.6cm] \displaystyle
= 4 \Tr [ (1 \mp {\gamma}^5) \hat{a}_1 \hat{a}_2 \cdots \hat{a}_{2n+1}
 \hat{b}_{2m+1} \cdots \hat{b}_2 \hat{b}_1 ] \; ,
\end{array}
\label{e.1-16}
\end{equation}
\begin{equation}
\begin{array}{l} \displaystyle
 \Tr ( {\gamma}_{\rho} \hat{a}_1 \hat{a}_2 \cdots \hat{a}_{2n+1} )
 \cdot
 \Tr ( {\gamma}^{\rho} \hat{b}_1 \hat{b}_2 \cdots \hat{b}_{2m+1} )
= 2 \Tr \left[ \hat{a}_1 \hat{a}_2 \cdots \hat{a}_{2n+1} \
 ( \hat{b}_1 \hat{b}_2 \cdots \hat{b}_{2m+1} +  \hat{b}_{2m+1} \cdots \hat{b}_2 \hat{b}_1 ) \right] \; .
\end{array}
\label{e.1-17}
\end{equation}
%


\section{Trace calculation of  Dirac $\gamma$-matrices
contracted with massless vectors}

\subsection{Application of massless vectors}

In the method of trace calculation, what was briefly explained in
Section 2, vectors  $a_i$ contracted with Dirac $\gamma$-matrices
are arbitrary. But there is a significant simplification  of
expressions for traces when vectors $a_i$ are massless (see e.g.
\cite{r6}, \cite{r7}).

For massless vectors the particular equation is true:
\begin{equation}
\displaystyle
 a_i^2 = (a_i l_0)^2 - (a_i l_1)^2 - (a_i l_2)^2 - (a_i l_3)^2 = 0 \; ,
\label{e.2-1}
\end{equation}
i.e.
\begin{equation}
\displaystyle
(a_i l_0)^2 - (a_i l_1)^2 = (a_i l_2)^2 + (a_i l_3)^2  \; ,
\label{e.2-2}
\end{equation}
or
\begin{equation}
\displaystyle
(a_i q_{+}) (a_i q_{-}) = (a_i e_{+}) (a_i e_{-})  \; .
\label{e.2-3}
\end{equation}

Thus the functions $F_k (a_i, a_j)$ become to be
like the function $F_1(a_i,a_j)$ multiplied by some factor:
\begin{equation}
\begin{array}{l} \displaystyle
F_2 (a_i, a_j) = 2 [ (a_i e_{+}) (a_j q_{-}) - (a_i q_{-}) (a_j
e_{+}) ] =
           \\[0.5cm] \displaystyle
= 2 [ (a_i e_{+}) {(a_j e_{+}) (a_j e_{-}) \over (a_j q_{+})} - (a_i q_{-}) (a_j e_{+})
{(a_j q_{+}) \over (a_j q_{+})} ]
 = - {(a_j e_{+}) \over (a_j q_{+})} 2 [ (a_i q_{-}) (a_j q_{+}) - (a_i e_{+}) (a_j e_{-}) ] \; ,
\end{array}
\label{e.2-4}
\end{equation}
that is
\begin{equation}
\displaystyle
F_2 (a_i, a_j) = - {(a_j e_{+}) \over (a_j q_{+})} F_1 (a_i, a_j) \; .
\label{e.2-5}
\end{equation}

In the similar way
\begin{equation}
\begin{array}{r} \displaystyle
F_3 (a_i, a_j) = - {(a_i e_{-}) \over (a_i q_{-})} \cdot {(a_j
e_{+}) \over (a_j q_{+})} F_1 (a_i, a_j)
 = {(a_i e_{-}) \over (a_i q_{-})} F_2 (a_i, a_j)
 = - {(a_j e_{+}) \over (a_j q_{+})} F_4 (a_i, a_j)
 \; .
\end{array}
\label{e.2-6}
\end{equation}
\begin{equation}
\displaystyle
F_4 (a_i, a_j) =  {(a_i e_{-}) \over (a_i q_{-})} F_1 (a_i, a_j) \; .
\label{e.2-7}
\end{equation}

Formulae of the traces become to be simpler by far.

Through the identity
\begin{equation}
\displaystyle
(1 \pm \gamma^5) \hat{q} Q (1 \pm \gamma^5) \hat{q}
 = \Tr [ (1 \pm \gamma^5) \hat{q} Q ] \ (1 \pm \gamma^5) \hat{q}
 \; ,
\label{e.2-8}
\end{equation}
which is valid for any massless vector $q$ and any operator $Q$, one can obtain
\begin{equation}
\begin{array}{l} \displaystyle
 {1 \over 2} \Tr [ (1 - \gamma^5) \hat{a}_1 \hat{a}_2 \hat{a}_3 \cdots \hat{a}_{2n} ]
  = {1 \over 4 (a_1 q_{-})} \Tr [ (1 + \gamma^5) \hat{q}_{-} \hat{a}_1 \hat{a}_2 \hat{a}_3
 \cdots \hat{a}_{2n} \hat{a}_1] =
          \\[0.5cm] \displaystyle
= {1 \over 4 (a_1 q_{-})} \cdot {1 \over 4 (a_3 q_{-})}
 \cdot
 \Tr [ (1 + \gamma^5) \hat{q}_{-} \hat{a}_1 \hat{a}_2 \hat{a}_3
       (1 + \gamma^5) \hat{q}_{-} \hat{a}_3
 \cdots \hat{a}_{2n} \hat{a}_1] =
          \\[0.5cm] \displaystyle
 = { \Tr [ (1 + \gamma^5) \hat{q}_{-} \hat{a}_1 \hat{a}_2 \hat{a}_3 ] \over 4 (a_1 q_{-})}
 \cdot
 {1 \over 4 (a_3 q_{-})} \Tr [ (1 + \gamma^5) \hat{q}_{-} \hat{a}_3
 \cdots \hat{a}_{2n} \hat{a}_1] = \cdots =
          \\[0.5cm] \displaystyle
 = { \Tr [ (1 + \gamma^5) \hat{q}_{-} \hat{a}_1 \hat{a}_2 \hat{a}_3 ] \over 4 (a_1 q_{-})}
\cdot   { \Tr [ (1 + \gamma^5) \hat{q}_{-} \hat{a}_3 \hat{a}_4 \hat{a}_5 ] \over 4 (a_3 q_{-})} \cdot
  \ \cdots \ \cdot
 { \Tr [ (1 + \gamma^5) \hat{q}_{-} \hat{a}_{2n-1} \hat{a}_{2n} \hat{a}_1 ] \over 4 (a_{2n-1} q_{-})}
 \; .
\end{array}
\label{e.2-9}
\end{equation}
Further
\begin{equation}
\begin{array}{l} \displaystyle
{1 \over 4 (a_i q_{-})} \Tr [ (1 + \gamma^5) \hat{q}_{-} \hat{a}_i \hat{a}_j \hat{a}_k ]
 = {1 \over 16 (a_i q_{-}) (a_j q_{+})}
 \Tr [ (1 + \gamma^5) \hat{q}_{-} \hat{a}_i \hat{a}_j (1 - \gamma^5) \hat{q}_{+} \hat{a}_j
 \hat{a}_k ] =
          \\[0.5cm] \displaystyle
 = {1 \over 32 (a_i q_{-}) (a_j q_{+})}
\cdot
 \Tr [ (1 + \gamma^5) \hat{q}_{-} \hat{a}_i \hat{a}_j \hat{q}_{+}
 (1 + \gamma^5) \hat{q}_{-} \hat{q}_{+} \hat{a}_j  \hat{a}_k ] =
          \\[0.5cm] \displaystyle
 = {1 \over 32 (a_i q_{-}) (a_j q_{+})}
 \cdot   \Tr [ (1 - \gamma^5) \hat{q}_{+} \hat{q}_{-} \hat{a}_i \hat{a}_j ]
   \Tr [ (1 + \gamma^5) \hat{q}_{-} \hat{q}_{+} \hat{a}_j \hat{a}_k ]
 = {F_1 (a_i, a_j) F_3^{\ast} (a_j, a_k) \over 2 (a_i q_{-}) (a_j q_{+})}
 \; ,
\end{array}
\label{e.2-10}
\end{equation}
and finally from (\ref{e.2-9}), (\ref{e.2-10}) it follows that
\begin{equation}
\begin{array}{l} \displaystyle
{1 \over 2} \Tr [ (1 - \gamma^5) \hat{a}_1 \hat{a}_2 \hat{a}_3 \cdots \hat{a}_{2n} ]
 = { F_1 (a_1, a_2) F_3^{\ast} (a_2, a_3) \cdots F_1 (a_{2n-1}, a_{2n})
 F_3^{\ast} (a_{2n}, a_1)
 \over 2^n (a_1 q_{-}) (a_2 q_{+}) (a_3 q_{-}) \cdots (a_{2n} q_{+}) }
  \; .
\end{array}
\label{e.2-11}
\end{equation}

From (\ref{e.2-5}), (\ref{e.2-6}) we have
\begin{equation}
\displaystyle
     { F_1 (a_i, a_j) F_3^{\ast} (a_j, a_k) \over (a_j q_{+}) }
 = - { F_2 (a_i, a_j) F_2^{\ast} (a_j, a_k) \over (a_j q_{-}) } \; ,
\label{e.2-12}
\end{equation}
and (\ref{e.2-11}) takes the form
\begin{equation}
\begin{array}{l} \displaystyle
{1 \over 2} \Tr [ (1 - \gamma^5) \hat{a}_1 \hat{a}_2 \hat{a}_3 \cdots \hat{a}_{2n} ]
 = (-1)^n
 \cdot { F_2 (a_1, a_2) F_2^{\ast} (a_2, a_3)
 \cdots F_2 (a_{2n-1}, a_{2n}) F_2^{\ast} (a_{2n}, a_1)
 \over 2^n (a_1 q_{-}) (a_2 q_{-}) (a_3 q_{-}) \cdots (a_{2n} q_{-}) }
  \; .
\end{array}
\label{e.2-13}
\end{equation}

Note that
\begin{equation}
\begin{array}{l} \displaystyle
 \Tr [ (1 - \gamma^5) \hat{a}_1 \hat{a}_2 \hat{a}_3 \cdots \hat{a}_{2n} ]
 =  \Tr [ (1 + \gamma^5) \hat{a}_2 \hat{a}_3 \cdots \hat{a}_{2n} \hat{a}_1 ]
= \left(  \Tr [ (1 - \gamma^5) \hat{a}_2 \hat{a}_3 \cdots
\hat{a}_{2n} \hat{a}_1 ]
 \right)^{\ast} \; ,
\end{array}
\label{e.2-14}
\end{equation}
then (\ref{e.2-11}) leads to the one more expression of traces
\begin{equation}
\begin{array}{l} \displaystyle
{1 \over 2} \Tr [ (1 - \gamma^5) \hat{a}_1 \hat{a}_2 \hat{a}_3 \cdots \hat{a}_{2n} ]
 = { F_3 (a_1, a_2) F_1^{\ast} (a_2, a_3)  \cdots F_3 (a_{2n-1}, a_{2n})
 F_1^{\ast} (a_{2n}, a_1)
 \over 2^n (a_1 q_{+}) (a_2 q_{-}) (a_3 q_{+}) \cdots (a_{2n} q_{-}) }
  \; .
\end{array}
\label{e.2-15}
\end{equation}

At last an identity [see (\ref{e.2-6}), (\ref{e.2-7})]
\begin{equation}
\displaystyle
     { F_3 (a_i, a_j) F_1^{\ast} (a_j, a_k) \over (a_j q_{-}) }
 = - { F_4 (a_i, a_j) F_4^{\ast} (a_j, a_k) \over (a_j q_{+}) } \; ,
\label{e.2-16}
\end{equation}
provide for
\begin{equation}
\begin{array}{l} \displaystyle
{1 \over 2} \Tr [ (1 - \gamma^5) \hat{a}_1 \hat{a}_2 \hat{a}_3 \cdots \hat{a}_{2n} ]
 = (-1)^n
 \cdot { F_4 (a_1, a_2) F_4^{\ast} (a_2, a_3)
 \cdots F_4 (a_{2n-1}, a_{2n})
 F_4^{\ast} (a_{2n}, a_1)
 \over 2^n (a_1 q_{+}) (a_2 q_{+}) (a_3 q_{+}) \cdots (a_{2n} q_{+}) }
  \; .
\end{array}
\label{e.2-17}
\end{equation}

The expressions for traces (\ref{e.2-11}), (\ref{e.2-13}),
(\ref{e.2-15}), (\ref{e.2-17}) are equivalent. Since in the case
of massless vectors
\begin{equation}
\begin{array}{l} \displaystyle
\left| F_1 (a_i, a_j) \right| = 2 \sqrt{ (a_i a_j) (a_i q_{-}) (a_j q_{+}) } \; ,
          \\[0.5cm] \displaystyle
\left| F_3 (a_i, a_j) \right| = 2 \sqrt{ (a_i a_j) (a_i q_{+}) (a_j q_{-}) } \; ,
          \\[0.5cm] \displaystyle
\left| F_2 (a_i, a_j) \right| = 2 \sqrt{ (a_i a_j) (a_i q_{-}) (a_j q_{-}) } \; ,
          \\[0.5cm] \displaystyle
\left| F_4 (a_i, a_j) \right| = 2 \sqrt{ (a_i a_j) (a_i q_{+})
(a_j q_{+}) } \; ,
\end{array}
\label{e.2-18}
\end{equation}
then
\begin{equation}
\displaystyle
 \left|
 {1 \over 2} \Tr [ (1 - \gamma^5) \hat{a}_1 \hat{a}_2 \hat{a}_3 \cdots \hat{a}_{2n} ]
 \right|
 = 2^n \sqrt{ (a_1 a_2) (a_2 a_3) \cdots (a_{2n} a_1) } \; ,
\label{e.2-19}
\end{equation}
(see also \cite {r6}).

\subsection{Calculation of traces in case of appearance the type ${0 \over 0}$ uncertainty}

The appearance of uncertainty type $\displaystyle {0 \over 0}$ is
possible during numerical calculation by reason of denominators
presence in the formulae (\ref{e.2-11}), (\ref{e.2-13}),
(\ref{e.2-15}), (\ref{e.2-17}).

For example, when the formula (\ref{e.2-17}) is being used and
orthonormal basis is chosen in according to (\ref{e.1-10}), the
uncertainty will appear, if
$$ \displaystyle
a_i = const \cdot  q_{+} $$
i.e. if
$$ \displaystyle
(a_i)_0 =  (a_i)_x \; . $$

The simplest way to avoid the appearing of uncertainties in such
points of phase space is to use for calculation here another basis
vectors, because obtained formulas are correct for any orthonormal
basis and numerical results received are independent from it's
choice. But such approach results in unnecessary complicating of
the computer program.

There are three solutions of this problem:

{\bf 1.} At the points of phase space, where denominators are 0,
one should use common formulae from Section 2 (see e.g. \cite{r8}).

{\bf 2.} One can choose an arbitrary 4-vector $\displaystyle t$
normalized by condition $\displaystyle t^2=1$ to perform the
identical transformation of the initial expression:
\begin{equation}
\begin{array}{l} \displaystyle
  {1 \over 2} \Tr [ (1 - \gamma^5) \hat{a}_1 \hat{a}_2 \hat{a}_3 \cdots \hat{a}_{2n} ]
= {1 \over 2} \Tr [ (1 + \gamma^5) \hat{t} \hat{a}_1 \hat{t}
\hat{t} \hat{a}_2 \hat{t}
 \hat{t} \hat{a}_3 \hat{t} \cdots \hat{t} \hat{a}_{2n} \hat{t} ] =
          \\[0.5cm] \displaystyle
= {1 \over 2} \Tr [ (1 + \gamma^5) \hat{a}'_1 \hat{a}'_2 \hat{a}'_3 \cdots \hat{a}'_{2n} ]
 = {1 \over 2} \Tr [ (1 - \gamma^5) \hat{a}'_2 \hat{a}'_3 \cdots \hat{a}'_{2n} \hat{a}'_1 ]
 \; ,
\end{array}
\label{e.3-1}
\end{equation}
where
\begin{equation}
\displaystyle
a'_i = - a_i + 2 (a_i t) t \; .
\label{e.3-2}
\end{equation}

Thus
\begin{equation}
\displaystyle
(a'_i)^2 = (a_i)^2 - 4 (a_i t)^2 + 4 (a_i t)^2 t^2 =  (a_i)^2 = 0
\; ,
\label{e.3-3}
\end{equation}
so one can use the formulae from Section 3.1 to calculate the transformed expression.
At the same time the denominators take forms
\begin{equation}
\displaystyle
(a'_i q_{\pm}) = - (a_i q_{\pm}) + 2 (a_i t) (t q_{\pm}) \; ,
\label{e.3-4}
\end{equation}
where the second term in (\ref{e.3-4}) is positive.

{\bf 3.} It is possible to transform the assumption formula. Let
us suppose that (\ref{e.2-17}) is using for calculation  and for
$i = 2s$
$$ \displaystyle ( a_{2s} q_{+} ) = 0 \; .$$
In this situation one can replace [see (\ref{e.2-16})]
\begin{equation}
\displaystyle
 { F_4 (a_{2s-1}, a_{2s}) F_4^{\ast} (a_{2s}, a_{2s+1})
   \over (a_{2s} q_{+}) }
 \rightarrow
 - { F_3 (a_{2s-1}, a_{2s}) F_1^{\ast} (a_{2s}, a_{2s+1})
   \over (a_{2s} q_{-}) }
  \; .
\label{e.3-5}
\end{equation}

Notice that in this case
$$ \displaystyle ( a_{2s} q_{-} ) \neq 0 \; . $$
%


\section{Conclusion}

The simple and compact formulae are proposed to calculate traces
of Dirac $\gamma$-matrices contracted with massless vectors. These
formulae may be easily implemented as a simple yet efficient
computer algorithm.




\end{document}